\documentclass[a4,11pt]{report}
\usepackage[english]{babel} 
\usepackage[latin1]{inputenc}   
\RequirePackage{lineno}
\usepackage{graphicx}
\usepackage[normalem]{ulem}
\usepackage{amssymb}
\usepackage{amstext}
\usepackage{amsfonts}
\usepackage[latin1]{inputenc}
\usepackage[normalem]{ulem}
\usepackage{epstopdf}
\usepackage{subfigure}
\usepackage{color}
\usepackage[dvipsnames]{xcolor}
\usepackage[justification=centering]{caption}
\usepackage{hyperref}
\usepackage{float}
\usepackage{graphics}
\usepackage{subfigure}
\usepackage{graphicx}
\usepackage{epsfig}
\usepackage[centertags]{amsmath}
\usepackage{graphicx,indentfirst,amsmath,amsfonts,amssymb,amsthm,newlfont}
\usepackage{longtable}
\usepackage{cite}

\begin{document}

\title{Understanding MIDI: A Painless Tutorial on Midi Format}

\author{
 {\large  H. M. de Oliveira}\thanks{hmo@de.ufpe.br also \url{https://arxiv.org/a/deoliveira_h_1.atom}}\\
 {\small Departamento de Estat\'istica, Universidade Federal de Pernambuco, UFPE, Recife, PE} \\
{\large R. C. de Oliveira}\thanks{rcoliveira@uea.br} \\
{\small Depto de Engenharia da Computa\c{c}\~ao, Universidade Estadual do Amazonas, UEA, Manaus, AM}
}
 
\criartitulo

\begin{abstract}
{\bf Abstract}. A short overview demystifying the midi audio format is presented. The goal is to explain the file structure and how the instructions are used to produce a music signal, both in the case of monophonic signals as for polyphonic signals.

\noindent
{\bf keywords}. Midi files, Computer Music, Monophonic and Polyphonic.
\end{abstract}

\section{Introduction}

Digital music is a broad and fascinating subject \cite{Arnell}, \cite{Cataltepe}, \cite{Pohlmann}. The Musical Instrument Digital Interface (MIDI) protocol is an industry-standard defined in early 80's to represent musical information \cite{Heckroth}, \cite{Matzkin}. Even dealing with an old subject, it does not seem to be available on the Internet enlightening texts (most shows only tables, commands and technicalities). It is a the most spread binary protocol for communicating intended to connect electronic musical instruments such as synthesizers, in computer for the purpose of recording, editing and programming, and electronic music equipment in the electronic home studios \cite{MIDI1}, \cite{Guerin}. MIDI System appeared for the need for standardization of the media between the synthesizers. This standard was created when electronic music was developed by a consortium of Japanese and American manufacturers Synthesizers (Sequential Systems, Roland Corporation, Yamaha, Kurzweil...). You can make free download millions of popular songs in MIDI versions. It transmits data using serial ports. For many people, still remains somewhat nebulous how to assemble a midi file from a music as well as the interpretation of a file previously generated. Unlike other formats (such as .wav and .mp3), a MIDI file does not contain audio itself, but the instructions to produce it, i.e. it is basically a digitized score. That is to say that MIDI is not audio signal: it just contains musical instruction. MIDI messages are comparable with those piano rolls, in which data control are represented as opposed to audio waveforms. The MIDI protocol allows sending messages over 16 independent channels, allowing 16 hearing instruments.  MIDI consists of both a simple hardware interface, and a more elaborate transmission protocol. What MIDI specifies? Here is a list: hardware; driver;  communication channels; messages; modes; controllers; files standard MIDI;  visual control (MIDI show control) among other things. In this note we are just concerned with midi protocol \cite{Cooper}, \cite{Heckroth}.

\section{Structure of the MIDI File}

\begin{figure}[ht]
 \centering
 \includegraphics[scale=0.23]{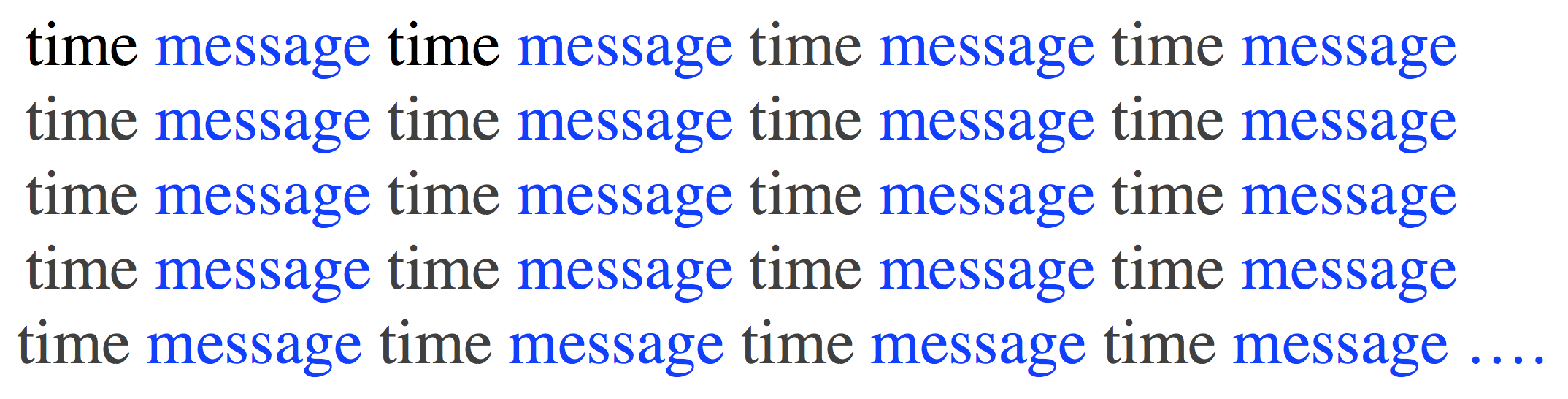}
 \caption{\small{Structure of a MIDI file.}}
 \label{fig:MIDI_STRUCTURE}
\end{figure}
Standard MIDI Files contain at the top level things called events. Each event consists of two components: a MIDI time, and a MIDI message. These {time/message} pairs follow each other one in a MIDI file as illustrated in Fig.~\ref{fig:MIDI_STRUCTURE}.
The time value is a measurement of the time to wait before playing the next message in the stream of MIDI file data. This method of specifying the time is called delta time that specifies the duration between two events (Table 1). MIDI commands and data are distinguished according to the most significant bit of the byte. If there is a zero in the top bit, then the byte is a data byte, otherwise the byte is a command byte.
\begin{table}[h]
\caption{ {\small 
{Understanding delta times in a sequence of events.}}}
\end{table}
\vspace{-1em}
\begin{figure}[ht]
 \centering
 \includegraphics[scale=0.4]{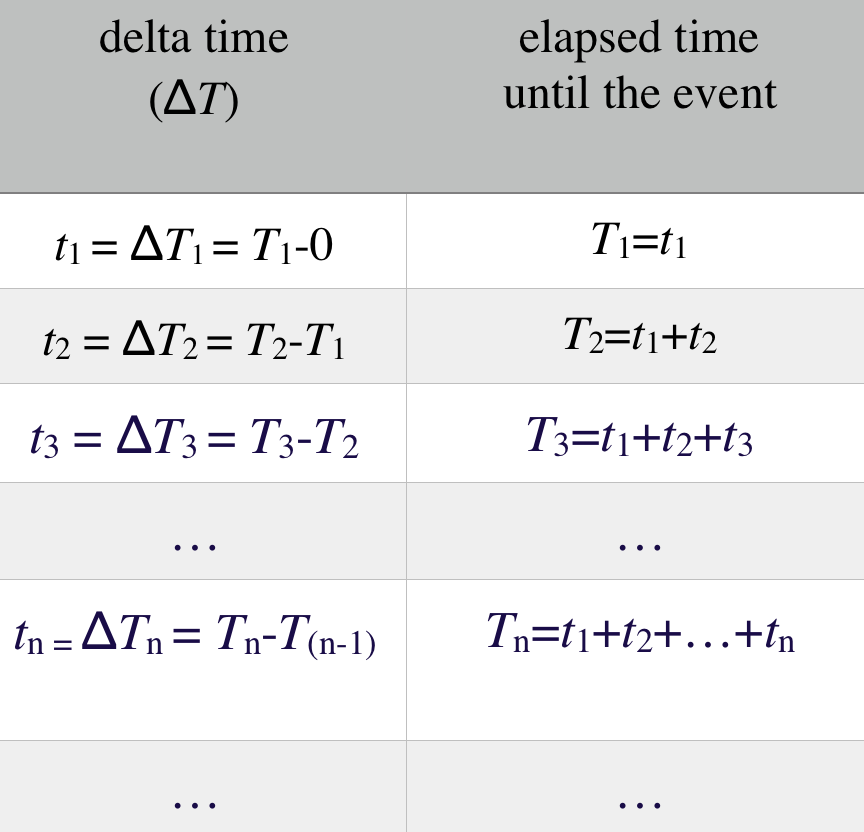}
 \end{figure}
\\The MIDI messages are sent as a sequence of one or more bytes. The first byte is a \textbf{\textcolor{blue}{command}} (STATUS) byte, often followed by \textcolor{blue}{data} (DATA) bytes with additional parameters. The command byte determines the type of the command. The number of DATA bytes that follow depend on the type of the message. For instance, the following two commands have different number of data bytes: \textbf{\textcolor{blue}{FF}} \textcolor{blue}{58 04 04 02 30 08} and \textbf{\textcolor{blue}{90}} \textcolor{blue}{3C 28}. The main commands in midi files are {Note-on} (0$\times$90) and {Note-off} (0$\times$80) that allows starting/stopping playing a single musical note. There are at least two playing modes, namely mono and poly. \textit{Monophonic}: the start of a new ``note-on'' command implies the termination of the previous note. \textit{Polyphonic}: multiple notes may be sounding at once, until the notes reach the end of their decay envelope, or when explicit ``note-off'' commands are received.

\section {About Midi Messages}
\begin{figure}[ht]
 \centering
 \includegraphics[scale=0.8]{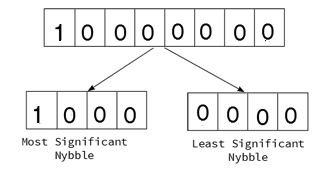}
 \caption{\small{Command type nibble (four bytes).}}
 \label{fig:nibble}
 \end{figure}

\noindent
The main messages are the NOTE ON and NOTE OFF messages. The NOTE ON message is sent when the performer hits a key of the music keyboard. It contains parameters to specify the pitch of the note as well as the velocity (i.e. intensity of the note when it is hit). When a synthesizer receives this message, it starts playing that note with the correct pitch and force level. When the NOTE OFF message is received, the corresponding note is switched off by the synthesizer. The channels are numbered 1-16, but their actual corresponding binary encoding is 0-15. MIDI commands can be further decomposed into a command type nibble (four bytes) and a channel number nibble (Fig.~\ref{fig:nibble}).
 \\
    command (hex 0$\times$80) $\Rightarrow$ channel nibble (1000  0000)\\
The NOTE ON (NOTE OFF, respectively) message is structured as follows:
\\
\indent
~~~~~~~~~~~~~~~~~~~~~~~~~note on ~~~~~~~~ note off\\
\indent
	command byte : ~~ 1001 cccc  ~~~~ 1000 cccc (MIDI channel, from 0 to 15)\\
\indent
	data byte 1 : ~~~~~~ 0ppp pppp ~~~~ 0ppp pppp (pitch value from 0 to 127)\\
\indent
	data byte 2 : ~~~~~~ 0vvv vvvv ~~~~ 0000 0000 (velocity/intensity value from 0 to 127)\\
For the midi sequence: 0$\times$90 0$\times$3c 0$\times$28, we have:
command: play a note in channel 0\\
\noindent
data: note is C4 (pitch, midi code 60=hex 0$\times$3c), its velocity ppp=20 (hex 0$\times$28).\\
\vspace{-2em}
\begin{table}[ht]
\centering
\begin{tabular}{c c | c c | c c}
\tabularnewline
\textbf{\textcolor{blue}{9}} & \textbf{\textcolor{blue}{0}} &  \textcolor{blue}{3} & \textcolor{blue}{c} & \textcolor{blue}{2} & \textcolor{blue}{8}\tabularnewline
\textbf{\textcolor{blue}{1001}} & \textbf{\textcolor{blue}{0000}} &  \textcolor{blue}{0011} & \textcolor{blue}{1011} & \textcolor{blue}{0010} & \textcolor{blue}{1000}\tabularnewline
\end{tabular}
\end{table}
\\The velocity value normally goes from 1 to 127, covering the range from a practically inaudible note up to the maximum note level. It basically corresponds to the scale of nuances found in music notation, as follows. Figure~\ref{fig:parameters} shows different standard note level. 
In the beginning of the file, it is quite common to use the following commands:
\begin{figure}[ht]
\centering
\includegraphics[scale=0.42]{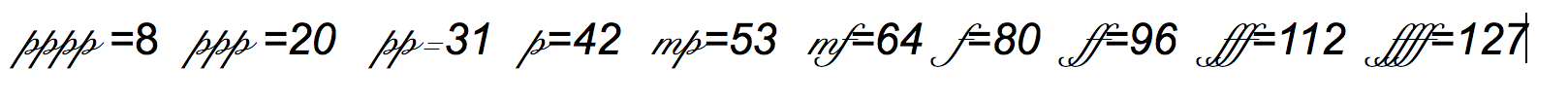}
\caption{\small{Parameter (data byte) associated with the note-on command.}}
 \label{fig:parameters}
\end{figure}
\\
midi time signature: FF 58 nn dd  cc bb (e.g. FF 58  4  4 128 18)\\
\indent
	nn  $\Rightarrow$ compass identification, e.g. for 4/4 set nn=4 dd=4\\
    \indent
	cc  $\Rightarrow$ number of ticks per metronome clock (reference note).\\
In this case, the ``reference note'' is 1/4 (crotchet), setting 128 ticks/crotchet, and the hex code associated with note value can be computed to state the delta time. \\
Another command used in the beginning encompasses the definition of the metronome: FF 51 03 tt tt tt set number microseconds per quarter note. For instance: FF 51 03 07 A1 20 for 120 bpm; FF 51 03 0f 42 40 for 60 bpm.\\
End of Track: FF 2F 00. This event is not optional. It is included so that an exact ending point may be specified for the track, so that an exact length is defined, which is necessary for tracks which are looped or concatenated. For looking inside a midi file, we suggest \cite{hex_editor}. For example, here is initial data from a monophonic MIDI file (in hex): \small{{00 FF 58 04 04 02 30 08 00 FF 59 02 00 00 00 90 3C 28 81 00 90 3C 00 00 90 3C 1E 81 00 90 3C 00 00 90 43 2D 81 00 90 43 00 00 90 43 32 81 00 90 43 00 00 90 45 2D 81 00 90 45 $\cdots$}} \normalsize{The way of reading this file is separating \{time/message\} pairs as} \small{\{00 \textcolor{blue}{ \textbf{FF} 58 04 04 02 30 08\}} \{00 \textcolor{blue} {\textbf{FF} 59 02 00 00\}} \{00 \textcolor{blue}{\textbf{90} 3C 28\}} \{81 00 \textcolor{blue}{\textbf{90} 3C 00\}} \{00 \textcolor{blue}{\textbf{90} 3C 1E\}} \{81 00 \textcolor{blue}{\textbf{90} 3C 00\}} \{00 \textcolor{blue}{\textbf{90} 43 2D\}} \{81 00 \textcolor{blue}{\textbf{90} 43 00\}} \{00 \textcolor{blue}{\textbf{90} 43 32\}} \{81 00 \textcolor{blue}{\textbf{90} 43 00\}} \{00 \textcolor{blue}{\textbf{90} 45 2D\}} \{81 00 \textcolor{blue}{\textbf{90} 45 00\}} $\cdots$} \normalsize{Black characters are delta time information, and blue characters are messages. Blue boldface type part is command and ordinary blue characters are data associated with the previous command. Final remark: many times the command note-off \textcolor{blue}{\textbf{80} nn 00} is replaced by the command \textcolor{blue}{\textbf{90} nn 00} (note-on, pitch nn, null intensity), which is equivalent.}

\section{Monophonic Audio on MIDI Files}
\begin{figure}[ht]
\centering
\includegraphics[scale=0.19]{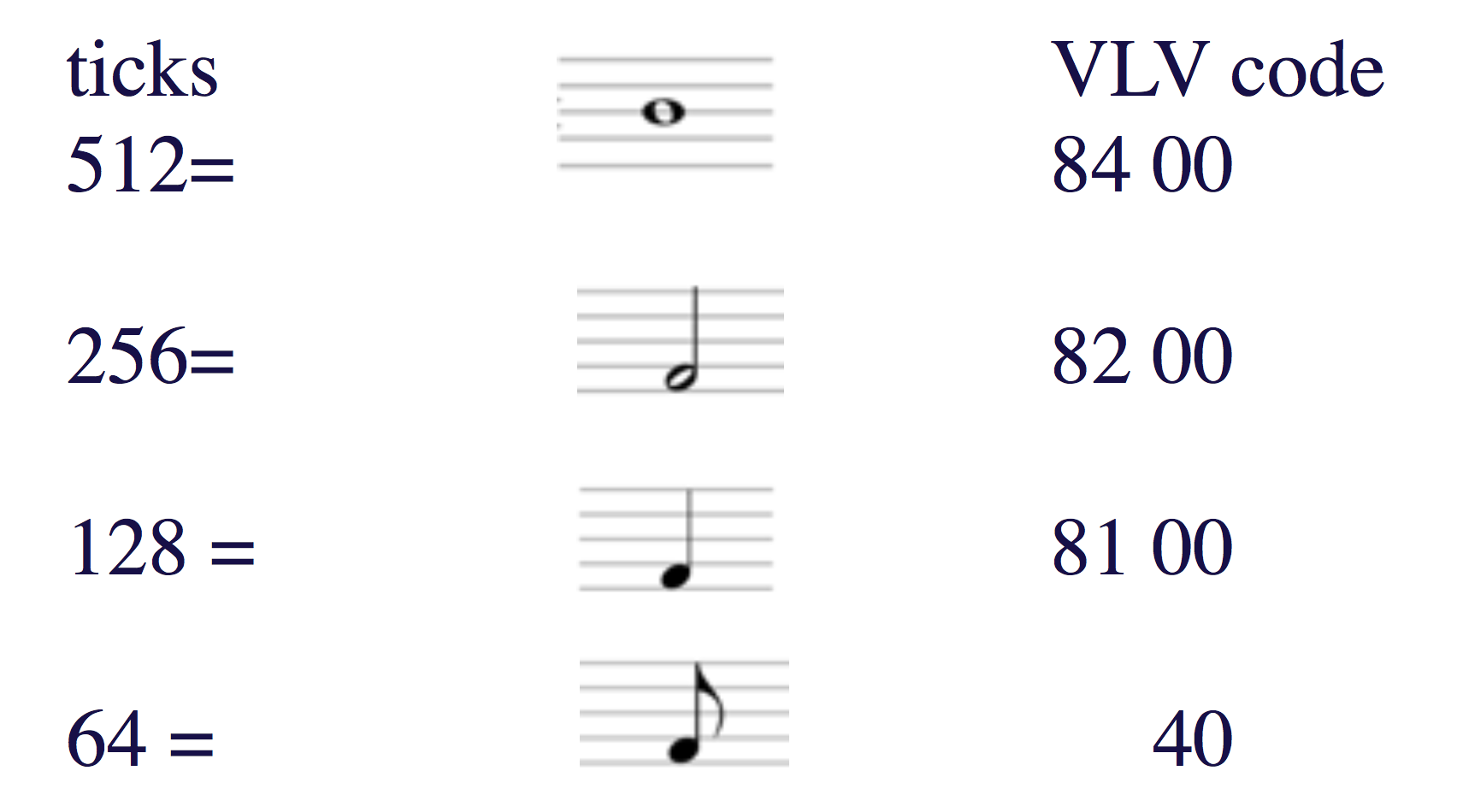}
\caption{\small{Number of ticks for the relative duration of a note.}}
 \label{fig:ticks}
\end{figure}

Time value are expressed in terms of the number of ticks. However, the hex-value to record the time value is not obtained just converting the number of ticks. A special format is used, which is referred to Variable Length Values (VLV, see \cite{Heckroth} for details). Notice that time values sometime are two bytes in length, and sometimes they are one byte in length. Using more than one byte for the delta time implies a longer time value. For example, here is some data from a simple monophonic MIDI file (in hex):\small{
00 \textcolor{blue}{\textbf{FF} 58 04 04 02 30 08} 
00 \textcolor{blue}{\textbf{FF} 59 02 00 00} 
00 \textcolor{blue}{\textbf{90} 3C 28} 
81 00 \textcolor{blue}{\textbf{90} 3C 00} 
00 \textcolor{blue}{\textbf{90} 3C 1E}
81 00 \textcolor{blue}{\textbf{90} 3C 00} 
00 \textcolor{blue}{\textbf{90} 43 2D}
81 00 \textcolor{blue}{\textbf{90} 43 00} 
00 \textcolor{blue}{\textbf{90} 43 32}
81 00 \textcolor{blue}{\textbf{90} 43 00} 
00 \textcolor{blue}{\textbf{90} 45 2D}
81 00 \textcolor{blue}{\textbf{90} 45 00} 
00 \textcolor{blue}{\textbf{90} 45 32}
81 00 \textcolor{blue}{\textbf{90} 45 00} 
00 \textcolor{blue}{\textbf{90} 43 23} 
82 00 \textcolor{blue}{\textbf{90} 43 00} 
00 \textcolor{blue}{\textbf{90} 41 32}
81 00 \textcolor{blue}{\textbf{90} 41 00} 
00 \textcolor{blue}{\textbf{90} 41 2D} 
81 00 \textcolor{blue}{\textbf{90} 41 00} 
00 \textcolor{blue}{\textbf{90} 40 32}
40 \textcolor{blue}{\textbf{90} 40 00}
40 \textcolor{blue}{\textbf{90} 40 28}
40 \textcolor{blue}{\textbf{90} 40 00}
40 \textcolor{blue}{\textbf{90} 3E 2D}
40 \textcolor{blue}{\textbf{90} 3E 00}
40 \textcolor{blue}{\textbf{90} 3E 32}
40 \textcolor{blue}{\textbf{90} 3E 00}
40 \textcolor{blue}{\textbf{90} 3C 1E} 82 00 \textcolor{blue}{\textbf{90} 3C 00} 
00 \textcolor{blue}{\textbf{FF} 2F 00}}
\normalsize{Do have then to look at the header of the MIDI file to understand what the units mean. For this example, the time units are 128 ticks to the quarter note, so 128 is a quarter note duration, 256 is a half-note, and 64 is an eighth-note duration (Fig.~\ref{fig:ticks}).}

\noindent
The sequence of played notes is \small{
C4 1/4, C4 1/4, G4 1/4,  G4 1/4, A4 1/4, A4 1/4,
G4 1/4, F4 1/4, F4	1/4, E4 1/8, E4 1/8, D4 1/8,
D4 1/8, C4 1/2}\normalsize{.} 

\section{Polyphonic Audio on MIDI}
\begin{figure}[H]
\center
\subfigure[score]{\includegraphics[scale=0.5]{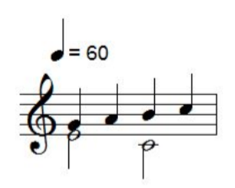}}
\qquad
\subfigure[colored score]{\includegraphics[scale=0.26]{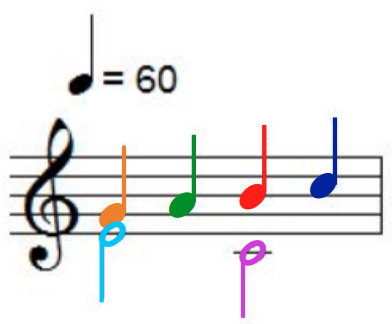}}
\captionsetup{justification=centering}
\caption{\small{Excerpt of a score.}}
\label{fig:polyscore}
\end{figure}

\begin{figure}[ht]
\centering
\includegraphics[scale=0.51]{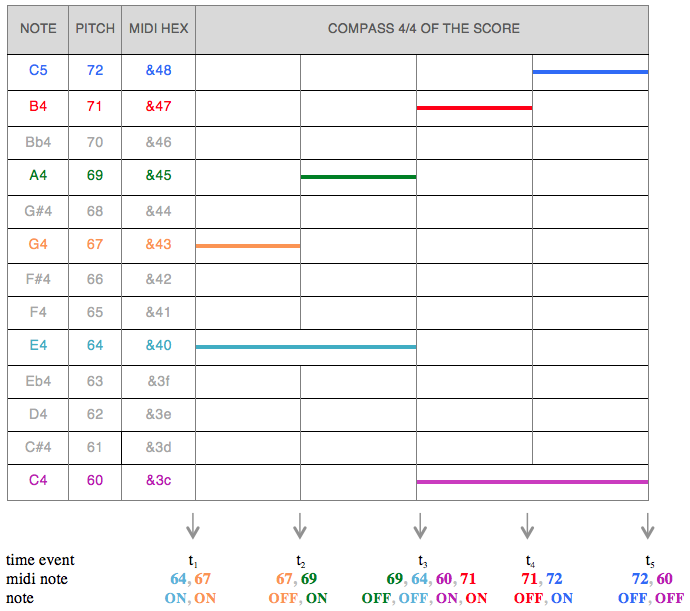}
\captionsetup{justification=centering}
\caption{\small{Excerpt of a score presented in Fig.~\ref{fig:polyscore}.}}
\label{fig:notegram}
\end{figure}

We are particularly interested in the midi perspective for musicians \cite{Airy_Parr}, \cite{Heckroth}. Let us take a straightforward example. It is easier to build the MIDI coding generating the corresponding notegram (Fig.~\ref{fig:notegram}) to excerpt of  score displayed in Fig.~\ref{fig:polyscore}. The list of midi events is \small{
00 \textcolor{blue}{\textbf{90} 40 40} 
00 \textcolor{blue}{\textbf{90} 43 40} 
81 00 \textcolor{blue}{\textbf{80} 43 00} 
00 \textcolor{blue}{\textbf{90} 45 40} 
81 00 \textcolor{blue}{\textbf{80} 45 00} 
00 \textcolor{blue}{\textbf{80} 40 00} 
00 \textcolor{blue}{\textbf{90} 3C 40} 
00 \textcolor{blue}{\textbf{90} 47 40} 
81 00 \textcolor{blue}{\textbf{80} 47 00} 
00 \textcolor{blue}{\textbf{90} 48 40} 
81 00 \textcolor{blue}{\textbf{80} 48 00 } 
00 \textcolor{blue}{\textbf{80} 3C 40}}.  

\small{\noindent
00 \textcolor{blue}{90 40 40}~~~~~~~(Start of \textbf{\textcolor{cyan}{E4 note}}, pitch = 64)\\
00 \textcolor{blue}{90 43 40}~~~~~~~(Start of \textbf{\textcolor{orange}{G4 note}}, pitch= 67)\\
81 00 \textcolor{blue}{80 43 00} ~~(End of \textbf{\textcolor{orange}{G4 note}}, pitch=67)\\
00 \textcolor{blue}{90 45 40}~~~~~~~(Start of \textbf{\textcolor{OliveGreen}{A4 note}}, pitch=69)\\
81 00 \textcolor{blue}{80 45 00} ~~(End of \textbf{\textcolor{OliveGreen}{A4 note}}, pitch=69)\\
00 \textcolor{blue}{80 40 00}~~~~~~~(End of \textbf{\textcolor{cyan}{E4 note}}, pitch=64)\\
00 \textcolor{blue}{90 3C 40}~~~~~~~(Start of \textbf{\textcolor{Plum}{C4 note}}, pitch = 60)\\
00 \textcolor{blue}{90 47 40}~~~~~~~(Start of \textbf{\textcolor{red}{B4 note}}, pitch= 71)\\
81 00 \textcolor{blue}{80 47 00}~~~(End of \textbf{\textcolor{red}{B4 note}}, pitch= 71)\\
00 \textcolor{blue}{90 48 40}~~~~~~~(Start of \textbf{\textcolor{blue}{C5 note}}, pitch= 72)\\
81 00 \textcolor{blue}{80 48 00}~~~(End of \textbf{\textcolor{blue}{C5 note}}, pitch= 72)\\
00 \textcolor{blue}{80 3C 40}~~~~~~~(End of \textbf{\textcolor{Plum}{C4 note}}, pitch = 60)}

\begin{figure}[ht]
\centering
\includegraphics[scale=0.33]{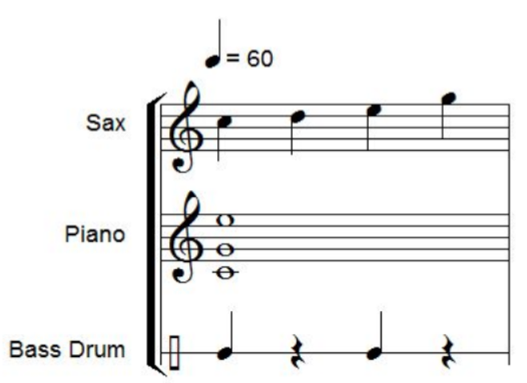}
\captionsetup{justification=centering}
\caption{\small{Playing different instruments: how midi works.}}
\label{fig:score_instruments}
\end{figure}
\normalsize
The MIDI message used to specify the instrument is a ``program change'' message: command byte=1100 CCCC (MIDI channel 0 to 15); Data byte 1=0XXX XXXX (instrument number from 0 to 127). The sax, piano and bass drum will use channels 1, 2 and 10, respectively. An illustrative example of MIDI sequence is shown in the sequel. First send the program changes to specify the instruments on each channel. In Fig. ~\ref{fig:score_instruments} there is a score example with three instruments. Suppose that the sax, piano and bass drum will use respectively channels 1, 2 and 10. The MIDI message sequence is the following. \\
\textbf{Specifying the instruments to use on each channel}: 00 C0 41 (Alto Sax=66 $\Rightarrow$ coded 65 =0$\times$41), 
	00 C1 00 (Piano=1 $\Rightarrow$ coded 0), 
	00 C9 00 (Standard Drums Kit=1 $\Rightarrow$ coded 0)\\
\textbf{Notes inserting}: 00 90 48 40 (Start sax C4, pitch=72=0$\times$48), 
	00 91 3C 40 (Start piano C3, pitch=60=0$\times$3C), 
	00 91 43 40 (Start piano G3, pitch=67=0$\times$43), 
	00 91 4C 40 (Start piano E4, pitch=76=0$\times$4C), 
	00 99 23 40 (Start Bass Drum=35=0$\times$23), 
	81 00 90 48 00 (Stop sax C4, pitch=72=0$\times$48), 
	00 99 23 00 (Stop Bass Drum=35=0$\times$23), 
	00 90 4A 40 (Start sax D4, pitch=74=0$\times$4A), 
	81 00 90 4A 00 (Stop sax D4, pitch=74=0$\times$4A), 
	00 90 4C 40 (Start sax E4, pitch=76= 0$\times$4C), 
	00 99 23 40 (Start Bass Drum =35=0$\times$23), 
	81 00 90 4C 00 (Stop sax E4, pitch=76=0$\times$4C), 
	00 99 23 00 (Stop Bass Drum=35=0$\times$23), 
	00 90 4F 40 (Start sax G4, pitch=79=0$\times$4F), 
	81 00 90 4F 00 (Stop sax G4, pitch=79=0$\times$4F), 
	00 91 3C 00 (Stop piano C3, pitch=60=0$\times$3C), 
	00 91 43 00 (Stop piano G3, pitch=67=0$\times$43), 
	00 91 4C 00 (Stop piano E4, pitch=76=0$\times$4C). Therefore, the midi sequence corresponding to the score of Figure ~\ref{fig:score_instruments} is \small{00 \textcolor{blue}{\textbf{C0} 41} 00 \textcolor{blue}{\textbf{C1} 00} 00 \textcolor{blue}{\textbf{C9} 00} 00 \textcolor{blue}{\textbf{90} 48 40} 00 \textcolor{blue}{\textbf{91} 3C 40} 00 \textcolor{blue}{\textbf{91} 43 40} 00 \textcolor{blue}{\textbf{91} 4C 40} 00 
\textcolor{blue}{\textbf{99} 23 40} 81 00 \textcolor{blue}{\textbf{90} 48 00} 00 \textcolor{blue}{\textbf{99} 23 00} 00 \textcolor{blue}{\textbf{90} 4A 40} 81 00 \textcolor{blue}{\textbf{90} 4A 00} 00 \textcolor{blue}{\textbf{90} 4C 40} 00 \textcolor{blue}{\textbf{99} 23 40} 81 00 \textcolor{blue}{\textbf{90} 4C 00} 00 \textcolor{blue}{\textbf{99} 23 00} 00 \textcolor{blue}{\textbf{90} 4F 40} 81 00 \textcolor{blue}{\textbf{90} 4F 00} 00 \textcolor{blue}{\textbf{91} 3C 00} 00 \textcolor{blue}{\textbf{91} 43 00} 00 \textcolor{blue}{\textbf{91} 4C 00}}\\

\section{Summary}
\normalsize
This brief explanatory report revises one of the representations of discrete mathematical structures used in music. It is intended for midi beginners. The motivation for writing it came from the difficulty finding a simple text  that allows understanding midi files. The presentation tries to throw some light in the way as a midi is built from a musical score. It also allows to understand the commands of a hex file (midi) generated to build the corresponding score. Fig.~\ref{fig:notegram} provides insight into the way the time events are defined. It is hoped that this note help ordinary users who deal with digital music. A final remark (just for readers who have never attempted to understand midi): ``when trying to learn midi by Internet, the reader could get a feel of the hardship in having clarity on the subject. Nevertheless, after reading this paper he may find it a trivial matter. Is that \textit{The mind that opens to a new idea never returns to its original size} (A. Einstein).''

\end{document}